\documentclass[twocolumn,showpacs,preprintnumbers,floatfix,amsmath,amssymb,prb]{revtex4}
\usepackage{graphicx}
\usepackage{amsmath}
\usepackage{dcolumn}
\usepackage{bm}
\newcommand{\qq}{{\mathbf q}}
\newcommand{\Hamil}{{\cal H}}
\newcommand{\kk}{{\mathbf k}}
\newcommand{\pp}{{\mathbf p}}
\newcommand{\RR}{{\mathbf R}}
\newcommand{\Tll}{$T_{\parallel}^{\mathrm{max}}$}
\newcommand{\Tperp}{$T_{\perp}^{\mathrm{max}}$}
\newcommand{\Tcoh}{$T^{\mathrm{coh}}$}

\begin{document}

\title{Coherence of polaronic transport in layered metals}

\author{Urban Lundin}
 \email{lundin@physics.uq.edu.au}
\author{Ross H.\ McKenzie}
\affiliation{Department of Physics, University of Queensland,
             Brisbane Qld 4072, Australia}
 
\date{\today}

\begin{abstract}
Layered systems shows anisotropic transport properties. 
The interlayer conductivity show a general temperature
dependence for a wide class of materials.
This can be understood if conduction occurs in two different channels
activated at different temperatures.
We show that the characteristic temperature dependence can be explained
using a polaron model for the transport. 
The results show an intuitive interpretation in terms
of coherent and incoherent quasi-particles within the layers.
Further, we extract results for the magnetoresistance, thermopower, 
spectral function and optical conductivity for the model and discuss 
application to experiments. 
\end{abstract}
\maketitle

\section{Introduction}
Layered materials show a range of interesting behavior, ranging from 
high temperature superconductivity to giant and colossal magneto-resistance. 
A common feature of some of these materials (see for example Ref.
\onlinecite{kimura96,hussey98,takenaka99}) is that they show a peak in 
the interlayer resistivity as a function of temperature. In some cases there 
is also a peak in the intralayer resistivity. 
We can identify different temperature scales, from experiment. 
\Tperp determines the maxima in the interlayer resistivity. 
\Tll determines the maxima in the intralayer resistivity. 
Besides this, recent angle resolved photoemission (ARPES) 
experiments~\cite{valla02} concluded that the peak in the interlayer 
resistivity is closely related to intralayer coherence, and that there 
is a crossover for the spectral function from being coherent to incoherent 
at a temperature \Tcoh. 
This idea that the scattering within the layers affect the transport between 
the layers have recently been investigated by Vozmediano 
{\textit et al.}~\cite{vozmediano03}. 
\begin{table}[h]
\caption{\label{expt:tab}
         Temperature scales in experiment for different materials. 
         n.s.\ means that the peak is not seen in the experiment, indicating 
         that, if it is there, it is higher than the temperature range 
         scanned in the experiment, n.a.\ means the result is not 
         available in our knowledge
}
\begin{ruledtabular}
\begin{tabular}{lccc} 
Material & \Tperp (K)  & \Tll (K)& \Tcoh (K)\\
\hline
(Bi$_{0.5}$Pb$_{0.5}$)$_2$Ba$_3$Cu$_2$O$_y$\footnotemark[1]
  & 200 & n.s.\ & $\sim$180 \\
NaCo$_2$O$_4$\footnotemark[1] 
  & 180 & n.s.\ & $\sim$150 \\
La$_{1.4}$Sr$_{1.6}$Mn$_2$O$_7$\footnotemark[2]
  & 100 & 270 & n.a.\ \\
Sr$_2$RuO$_4$\footnotemark[3]
  & 130 & n.s.\ & n.a.\ \\
TmBa$_2$Cu$_3$O$_{6.41}$\footnotemark[4] 
  & 127 & n.s.\ & n.a.\ \\ 
(TMTSF)$_2$PF$_6$\footnotemark[5]
  & 90 & n.s.\ & n.a.\ \\ 
$\kappa$-(BEDT-TTF)$_2$Cu(SCN)$_2$\footnotemark[6]
  & 95 & 100 & n.a.\ 
\end{tabular}
\end{ruledtabular}
\footnotetext[1]{from Ref. \onlinecite{valla02}}
\footnotetext[2]{from Ref. \onlinecite{kimura96}}
\footnotetext[3]{from Ref. \onlinecite{hussey98}}
\footnotetext[4]{from Ref. \onlinecite{lavrov98}}
\footnotetext[5]{from Ref. \onlinecite{mihaly00}}
\footnotetext[6]{from Ref. \onlinecite{buravov92}} 
\end{table}
All of this can be explained if there are two mechanisms of 
transport.~\cite{lavrov98,merino00} 
One, a coherent, dominates at low temperatures, while at elevated 
temperatures an incoherent contribution starts to dominate. 
In this paper we will demonstrate that polaronic transport can be the 
mechanism providing this physics. 
This gives an intuitive explanation for the different temperature scales 
associated with transport and coherence. 
We extend the idea presented by Alexandrov and Bratkovsky~\cite{alexandrov99}, 
and apply it to \textit{layered} transport. In that paper they discussed 
(bi)polaron formation within giant magnetoresistance materials. 

Another powerful tool when studying polaronic transport is to study the 
thermopower. At high temperatures the conductivity is activated and the 
resistivity shows an exponential temperature dependence with a gap 
$E_{\sigma}$, the thermopower usually shows a 
$1/T$-behavior where the barrier is $E_s$. One signature of polaronic 
transport is that $E_s<<E_{\sigma}$, whereas for normal semiconductors 
(where the transport is activated) we have $E_s=E_{\sigma}$~\cite{salamon01}. 
By comparing the high temperature electrical resistivity and thermopower 
a number of experimentalists have argued for the existence of small polarons 
in LaMnO$_3$-compounds~\cite{palstra97,liu00,chen03}. Further evidence for 
small polarons can be found in neutron scattering data, where the polaron 
induces a local deformation of the 
lattice~\cite{billinge96,adams00,campbell03}. 
Measurements of thermopower in different directions in an organic 
quasi-two-dimensional 
crystal found different behaviour between the interlayer thermopower and 
the interlayer one~\cite{choi02}. Further the presence of polarons was 
confirmed by photoemission experiments~\cite{mannella04}.

The approach we present is based on known approximations for the 
polarons~\cite{mahan,appel}. Recent dynamical mean field (DMFT) calculations 
made on the transport of small polarons~\cite{fratini03} indicate that the 
approximations we are going to use overestimates the resistivity, and 
the exact functional behavior of the resistivity. The results of 
Fratini {\textit et al.}~\cite{fratini03} shows that there are two 
temperature regions. 
One, semiconducting region where transport is heavily influenced by 
phonon fluctuations. Then a nonadiabatic regime, which compares mostly 
to the small polaron regime in the Holstein model.  
Here, however, we are more concerned with transitions between different 
regions 
of small polaron transport, not so much with the exact details, and it seems 
that the approximations we use does capture the essential physics. 
We do not claim that polarons are responsible for all the observed effects, 
simply that it can provide some insight into the physics of layered systems. 
For instance, for the manganites it seems that the double exchange model 
is the preferred one (see Ref.~\onlinecite{liu01} and references therein), 
although another explanation, in terms of a carrier density 
collapse due to bipolarons and their magnetic 
features~\cite{alexandrov99_2} is gaining interest. 
Even the thermopower seems to be consistent 
with this model~\cite{alexandrov01}. 
There are also theories using a combination of double exchange models and 
polarons for the localized structure~\cite{weisse03}. 
A shorter presentation of some of the result from this paper has been 
previously published~\cite{lundin03}. We have also investigated the 
problem of angular magnetoresistance oscillations in layered metals arising 
from incoherence~\cite{lundin04}. 

The layout of the paper is as follows, in section~\ref{model:sect}  
we present the model and the small polarons are introduced in 
section~\ref{small:sect} with the decay and Green function. 
In section~\ref{transport:sect} we turn to the transport properties for the 
intralayer and interlayer currents and thermopower. Last we conclude with 
a calculation of the optical properties, section~\ref{optical:sect}, and 
a special case of the magnetoresistance in section~\ref{MR:sect}. 

\section{Model Hamiltonian}
\label{model:sect}
\begin{figure}[hbt]
\includegraphics[width=\columnwidth]{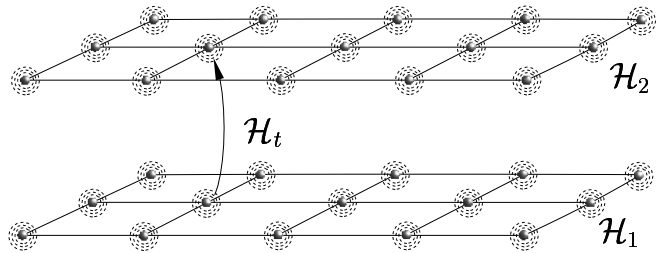}
\caption{\label{layers:fig}
         We model the two coupled layers as a anisotropic 3D system. 
         Within each layer the electrons couple to bosons to potentially 
         form small polarons. This is described by the the Hamiltonians 
         $\Hamil_1$ and $\Hamil_2$. 
         The two layers are the coupled by a direct hopping term, $\Hamil_t$.}
\end{figure}

We start with a Holstein model~\cite{holstein59} for an infinite system
where the electrons interact with bosons. 
The Hamiltonian is
\begin{eqnarray}
{\cal H}&=&\sum_i\epsilon^0c^{\dag}_ic_i+
           \sum_{\qq} \hbar\omega_{\qq}a^{\dag}_{\qq}a_{\qq}+
           \sum_{<i\eta>}t_{i\eta}c^{\dag}_{\eta}c_i 
\nonumber \\
&&
           + \sum_{i,\qq}M_{\qq}c^{\dag}_ic_ie^{i\qq\cdot\RR_i}
                (a_{\qq}+a^{\dag}_{-\qq}), 
\label{ham:eq}
\end{eqnarray}
where $\epsilon^0$ is the on-site energy, $\omega_{\qq}$ is the dispersion 
of the bosons, $t_{i\eta}$ is the hopping integral between neighboring  
sites $i$ and $\eta$, $M_{\qq}$ is the coupling between the bosons and 
the electrons. We want to emphasize that we will talk about bosons, since 
the theory will look the same for all types of bosons with a coupling 
given in the Hamiltonian above. The bosons can be phonons, spin-waves, 
plasmons, or any other type fulfilling bosonic commutation rules. 
Since we want to study layered systems we split the hopping
into parallel and perpendicular to the layers, 
$t_{\parallel}$ and $t_{\perp}$ respectively, 
where $t_{\parallel}\gg t_{\perp}$. 
We only include hopping between nearest neighbors, both for the intra- and 
interlayer hopping. 
This enables us to write the
Hamiltonian in a way more adapted for the layered case, shown
in Fig.~\ref{layers:fig}.
The nature of the transport depends on how $t_{\parallel}$ and $t_{\perp}$
compares with $\Gamma$, the scattering rate due to the bosons.
We assume that that $\Gamma > t_{\perp}$, so that the interlayer transport
can be described by considering two decoupled layers. 
Meaning that the electrons scatters many times within the layers before 
jumping to the next~\cite{vozmediano03,hussey98,lundin03,lundin04}. 
We assume that we can decouple the bosons within each layer separately, 
i.e., the bosons are localized in each layer, the only interaction between 
bosons in different layers come from the electron tunneling, but this event 
occurs more seldom than the electron scattering of bosons in each layer. 
The Hamiltonian can be specified for this system. 
Two layers are coupled with a hopping Hamiltonian. Within each layer
the electrons can hop but there is a coupling to a bosonic degree of freedom
in each layer. 
We then use the Hamiltonian:
\begin{equation}
{\cal H}={\cal H}_1+{\cal H}_2+{\cal H}_t \nonumber
\end{equation}
where
\begin{eqnarray}
{\cal H}_1&=&\sum_i\epsilon^0c^{\dag}_ic_i+
           \sum_{\qq}\hbar\omega_{\qq}a^{\dag}_{\qq}a_{\qq}+
           t_{\parallel}\sum_{<i\eta>}c^{\dag}_{\eta}c_i 
           + \sum_{i,\qq}M_{\qq}c^{\dag}_ic_ie^{i\qq\cdot\RR_i}
                (a_{\qq}+a^{\dag}_{-\qq}), \nonumber \\
{\cal H}_2&=&\sum_j\epsilon^0d^{\dag}_jd_j+
           \sum_{\pp}\hbar\omega_{\pp}a^{\dag}_{\pp}a_{\pp}+
           t_{\parallel}\sum_{<j\delta>}d^{\dag}_{\delta}d_j 
            + \sum_{j,\pp}M_{\pp}d^{\dag}_jd_je^{i\qq\cdot\RR_j}
                (b_{\pp}+b^{\dag}_{-\pp}), \nonumber \\
{\cal H}_t&=&t_{\perp}\sum_{i}(c^{\dag}_id_i+{\rm h.c.}). \nonumber
\end{eqnarray}
Here, and below, $\qq,c,i,a,1$ refers to one layer, and
$\pp,d,j,b,2$ to the other one. 

\section{Small polarons}
\label{small:sect}
First we focus on the properties of the two layers separately, i.e.,
we ignore the hopping term, ${\cal H}_t$ between the layers.
We perform a Lang-Firsov transformation~\cite{lang63}
to diagonalize the Hamiltonian, excluding the hopping term, defined above, 
in each layer.
Then $c_i \to \tilde{c}_i = c_i X_i$, 
and  $d_j \to \tilde{d}_j = d_j Y_j$ where
\begin{eqnarray}
&&X_i=\exp\left[\sum_{\qq}e^{i\qq\cdot\RR_i}\frac{M_{\qq}}{\hbar\omega_{\qq}}
          (a_{\qq}-a^{\dag}_{-\qq})\right] , \nonumber \\
&&Y_j=\exp\left[\sum_{\pp}e^{i\pp\cdot\RR_j}\frac{M_{\pp}}{\hbar\omega_{\pp}}
          (b_{\pp}-b^{\dag}_{-\pp})\right] 
\end{eqnarray}
are the polaron operators~\cite{mahan} for the first and second layer 
respectively. Further, 
$a_i \to a_i - {M \over \omega_0} c_i^\dagger c_i$.
The Hamiltonian is transformed to
$\bar{{\cal H}}=e^S{\cal H}e^{-S}$
where $S=\frac{M}{\hbar\omega_0}\sum_i c_i^\dagger c_i(a_i^\dagger -a_i)$.
The Hamiltonian becomes 
\begin{eqnarray}
\bar{{\cal H}}&=&
\sum_{\qq}\hbar\omega_{\qq}a^{\dag}_{\qq}a_{\qq}+
\sum_{\pp}\hbar\omega_{\pp}b^{\dag}_{\pp}b_{\pp}
-\sum_j\Delta d^{\dag}_jd_j
-\sum_i\Delta c^{\dag}_ic_i \nonumber \\
&&+ t_{\parallel}\sum_{i,\eta}(c^{\dag}_{i+\eta}c_iX^{\dag}_{i+\eta}X_i+
                               {\rm h.c.})
  + t_{\parallel}\sum_{j,\delta}(d^{\dag}_{j+\delta}d_jY^{\dag}_{j+\delta}Y_j+
                               {\rm h.c.}) \nonumber \\
&&+t_{\perp}\sum_{i,j}(c^{\dag}_id_jX^{\dag}_{i}Y_j+{\rm h.c.}),
\end{eqnarray}
where 
\begin{equation}
\Delta=\sum_{\qq}\frac{M_{\qq}^2}{\hbar\omega_{\qq}}-\epsilon^0, 
\end{equation}
is the polaron binding energy. 
The intralayer hopping terms can be treated by adding and subtracting
to the Hamiltonian a term~\cite{alexandrov} 
\begin{equation}
{\cal H}_{sp}\equiv
t_\parallel \sum_{<ij>}
 \langle X_i X^{\dag}_j\rangle
c^{\dag}_ic_j
\equiv \sum_{\kk} \epsilon_{\kk}
c^{\dag}_{\kk}c_{\kk}
\end{equation}
where $\langle  ..\rangle$ denotes a thermal average over boson states
and this term describes a tight-binding band of small polarons
for a square lattice within each layer~\cite{lang63,alexandrov}
\begin{equation}
\epsilon_{\kk} = \epsilon^0
- e^{-N^{-1}\sum_{\qq}\left(\frac{M_{\qq}}{\hbar\omega_{\qq}}\right)^2
(1+2n_B)}
t_\parallel [ \cos(k_x a  ) + \cos(k_y a ) ],
\label{tight:eq}
\end{equation}
where $a$ is the lattice constant within the layers, $N$ is the number 
of sites in one layer, and
$n_B(T) = (\exp( \hbar \omega_{\qq}/ k_B T) -1 )^{-1}$ is the Bose function.
We see that the quasi-particles are described by an tight binding energy, 
where the bandwidth is reduced due to the polaron formation. 
Polaron transport narrowing has been seen experimentally in 
muon-experiments.~\cite{karlsson} 

There is then a residual interaction~\cite{alexandrov}
between the polarons and the bosons which is described by
\begin{equation}
\bar{\Hamil}_{p-b}=
t_\parallel \sum_{<ij>}
 [X_i X^{\dag}_j - \langle X_i X^{\dag}_j\rangle ]
                       c^{\dag}_ic_j,
\label{interaction:eq}
\end{equation}
and leads to scattering of the small polarons.

\subsection{Decay of the quasi-particles}
\label{decay:sect}
\begin{figure}[hbt]
\includegraphics[width=\columnwidth]{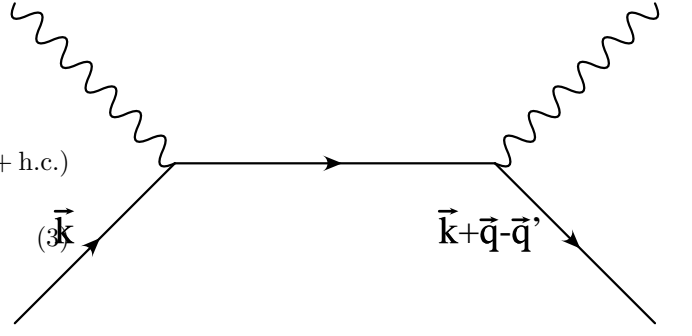}
\caption{\label{selfenergy:fig}
         Diagram describing the first contribution to the polaron decay, 
      $\Gamma$. The polaron emits and absorbs one boson, changing its momenta. 
}
\end{figure}

Later we will need the decay, $\Gamma$, so we start by calculating it. 
We will calculate the first contribution to the self energy 
in one layer by a method similar to the one used 
by Alexandrov and Mott~\cite{alexandrov}. 
The first non-zero contribution to the imaginary part of the self-energy,
$\Sigma$, comes 
when the polaron emits one boson and absorb one boson. This process is shown 
in Fig.~\ref{selfenergy:fig}, and is induced by the polaron-boson scattering 
from Eq.~(\ref{interaction:eq}). 
By using Fermi's Golden rule we get an expression for the decay:
\begin{eqnarray}
\Gamma=&&2\pi\sum_{\qq,\qq'}\left|\langle n_{\qq}-1,n_{\qq'+1};\kk+\qq-\qq'|
                             \Hamil_{p-b}|n_{\qq},n_{\qq'};\kk\rangle\right|^2
\nonumber \\
&&\times
                     \delta(\epsilon_{\kk}-\epsilon_{\kk+\qq-\qq'}).
\end{eqnarray}
Here, $\bar{\Hamil}_{p-b}$ is the polaron-boson interaction in 
Eq.~(\ref{interaction:eq}).  Using this Hamiltonian we get that
\begin{eqnarray}
&&\langle n_{\qq}-1,n_{\qq'+1};\kk+\qq-\qq'|
       \bar{\Hamil}_{p-ph}|n_{\qq},n_{\qq'};\kk\rangle
=
\nonumber \\ 
&&\frac{4t_{\parallel}}{N}\sqrt{n_{\qq}}\sqrt{n_{\qq'}+1}
      \left(\frac{M_{\qq}}{\hbar\omega_{\qq}}\right)
      \left(\frac{M_{\qq'}}{\hbar\omega_{\qq'}}\right)
 \langle \kk+\qq-\qq'|c^{\dag}_{\kk+\qq-\qq'}c_{\kk}|\kk\rangle 
\nonumber \\ &&\times
 \delta_{\qq-\qq'}.
\end{eqnarray}
To simplify this to get a energy independent expression we 
use a energy independent density of states and assume a $\kk$-independent 
coupling between the electrons and the bosons. 
We only consider a single frequency $\omega_0$ for reasons of simplicity;
it allows us to express some of our results in an analytical form. 
Then we can define the dimensionless coupling 
\begin{equation}
g\equiv \left(\frac{M}{\hbar\omega_0}\right)^2 \nonumber 
\end{equation}
that will enter our equations later. 
We require that $g\agt 1$ in order for small polaronic effects to be 
important.\footnote{Strictly speaking there are three conditions for small 
polaron transport~\cite{holstein59}, 
$t<g\hbar\omega_0$, 
$\sqrt{2}t<M$, and 
$t<\sqrt{M}\left(\frac{2k_BT\hbar\omega_0}{\pi^3}\right)^{1/4}$.}
Using this, we get that 
\begin{equation}
\tau^{-1}=Wg^2n_B(1+n_B),
\label{decay:eq}
\end{equation} 
where $W=4\tilde{t}_{\parallel}$ 
is the polaron bandwidth, which is subject to narrowing due to the 
renormalization of the hopping 
$t\rightarrow \tilde{t}_{\parallel} \equiv t_{\parallel} e^{-g(1+2n_B)}$. 

\subsection{Green function in the layer}
\label{GF:sect}
\begin{figure}[hbt]
\includegraphics[width=\columnwidth]{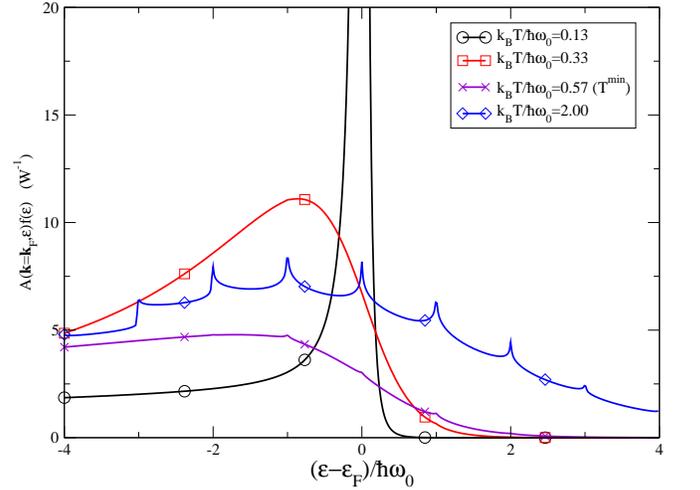}
\caption{\label{spectral:fig} 
         The quasi-particle spectral function, 
         $n_f(\epsilon)\mathrm{Im}[G(\kk_F,\epsilon)]$
         for a electron-phonon coupling, $g=1$, 
         for different temperatures. The sum over $\kk$ is done by 
         integrating over the density of states, which we assume is flat with 
         a bandwidth $W=77 \hbar\omega_0$. There are two contributions to the 
	 spectra function, one coherent dominating at low temperatures, and 
	 one incoherent dominating at high temperatures. Similar 
 	 behavior have been seen experimentally~\cite{valla02}.}
\end{figure}

Let us start by calculating the {\em electron} Green function (GF) within one 
layer, ignoring the coupling between the layers ($t_{\perp}=0$). 
This gives us valuable information on coherence of the quasi-particles, and 
can be compared to angle resolved photo-emission spectra (ARPES). 
After performing the 
Lang-Firsov transformation the small polaron GF is
\begin{eqnarray}
G^0(\kk,\tau)&=&-i\Theta(\tau)\frac{1}{N}\sum_{i,i'}
                 e^{i\kk\cdot(\RR_i-\RR_{i'})}
                \langle T_{\tau} c_i(\tau)c^{\dag}_{i'}(0)\rangle \nonumber \\
&=&-i\Theta(\tau)e^{(\epsilon_{\kk}-i\Gamma) i\tau /h}. \nonumber 
\end{eqnarray}
To get the electron GF we have to convolute this GF with the average over two 
polaron operators 
$\langle T X^{\dag}_i(t)X_{i'}(0)\rangle\equiv\Phi_{ii'}(t)$. 
This average can be decoupled and written as an 
exponential,~\cite{lang63,mahan}
\begin{eqnarray}
&&\Phi_{ii'}(t)=
e^{-g
                    (1/2+n_B)}
\nonumber \\
&&\times
\exp\left\{g\sum_{\qq}
                   \cos[\qq\cdot(\mathbf{R}_i-\mathbf{R}_{i'})]
                   [(1+n_B)e^{-i\omega t} +n_Be^{i\omega t}]\right\}. 
\nonumber \\
\label{expo:eq}
\end{eqnarray}
After Fourier transforming the average of the polaron operators, 
giving a sum of delta functions, 
we will have a convolution
\begin{equation}
G(\kk,i\omega_n)=
\frac{1}{N}\sum_{\omega_{n'},\RR_m,\kk'}\Phi(\RR_m,\omega_{n'}-\omega_n)
           G^0(\kk',\omega_{n'})e^{i(\kk-\kk')\cdot \RR_m}. \nonumber
\end{equation}
After some algebra we come to the following expression 
\begin{eqnarray}
G(\kk,i\omega_n)&=&
e^{-g(1+2n_B)}\frac{1}{N}
\sum_{\RR_m,\kk'}e^{i(\kk-\kk')\cdot\RR_m}
\nonumber \\ &&\times
\sum_{l=-\infty}^{\infty} 
\frac{I_l[2g\sum_{\qq}\cos(\qq\cdot\RR_m)
           \sqrt{n_B(1+n_B)}]e^{-l\hbar\omega_0\beta/2}}
     {i\omega_n-\epsilon_{\kk'}+l\hbar\omega_0+i\Gamma}. \nonumber \\
\end{eqnarray}
Here $I_l$ indicates a modified Bessel function of order $l$. 
Performing the summation over $\RR_m$, care has to be taken when considering
the $l=0$ term,  we get the final result for the GF 
\begin{eqnarray}
&&G(\kk,\omega)=e^{-g(1+2n_B)}
\left\{
       \frac{1}{\omega-\epsilon_{\kk}+i\Gamma} \right. \nonumber \\
&&+
       \sum_{\kk'}\frac{I_0\left[2g\sqrt{n_B(1+n_B)}\right]-1}
                       {\omega-\epsilon_{\kk'}+i\Gamma} \nonumber \\
&&\left.+\sum_{\kk',l\neq 0}
            \frac{I_{l}\left[2g\sqrt{n_B(1+n_B)}\right]
                  e^{-l\hbar\omega_0\beta/2}}
                 {\omega-\epsilon_{\kk'}+l\hbar\omega_0+i\Gamma}
\right\}. 
\label{GF:eq}
\end{eqnarray}
Note that we have written the GF as a sum of a coherent and an incoherent 
part. This can be compared to the zero temperature result by Alexandrov and 
Mott~\cite{alexandrov}. At $T=0$ there are no bosons to absorb and only 
$l\geq 0$ contributes to the GF. Also, we can 
compare this to the nonzero temperature GF by Ciuchi 
{\textit et al.}~\cite{ciuchi97}. 
The first line is dependent on $\kk$, thereby describing a coherent part. 
There will be a well-defined quasiparticle peak at $\omega=\epsilon_{\kk}$, 
with a spectral weight of $e^{-g(1+2n_B)}$. 
The second and third lines contains a sum over intralayer momentum 
and are therefore independent of $\kk$, they are incoherent. 
The two contributions have different temperature dependence, 
the coherent dominates at low temperature, and the incoherent at high 
temperature. 
This means that there is a crossover from coherent intralayer motion at low 
temperature to incoherent intralayer motion at high temperatures. 
In Fig.~\ref{spectral:fig} we show the spectral function resulting from this 
GF at the Fermi wave-vector as a function of energy. 
The sum over $\kk$ is done by integrating over a flat density of states. 
This is what is measured in ARPES experiments like the one in 
reference \onlinecite{valla02} for 
(Bi$_{0.5}$Pb$_{0.5}$)$_2$Ba$_3$Co$_2$O$_y$ and NaCo$_2$O$_4$. 
Recently, similar features was seen in Sr$_2$RuO$_4$.~\cite{wang04} 
The coherent contribution display a peak at the Fermi energy and 
$\kk$-vector at low temperature, indicating a coherent quasi-particle. 
The peak disappears as the temperature is increased. 
From plots of the spectral function one can estimate a crossover 
temperature when the contribution from the incoherent starts to dominate over 
the incoherent one. This will take place when
\begin{equation}
k_BT^*\sim\frac{\hbar\omega_0}{2g}.
\end{equation}
Later we will see that this relates to the temperature dependence of the 
interlayer conductivity in a special way. 

\section{Transport properties}
\label{transport:sect}
Let us now turn to the transport properties of the layered material which 
is described by the layered Hamiltonian defined above. 
At an applied voltage $V$, the current is given by 
the current-current correlation function from the Kubo formula~\cite{mahan} 
\begin{equation}
I_{\mu\nu}(eV)=\frac{2e}{h}\mathrm{Im}\left\{\int_0^{\beta} dte^{ieVt}
\langle T \hat{j}_{\mu}(t)\hat{j}^{\dag}_{\nu}(0)\rangle\right\},
\label{kubo:eq}
\end{equation}
where $\hat{j}$ is the current operator. $\mu$ and $\nu$ are 
directions in the crystal. 
For our system, including polarons, the current operator 
for nearest neighbor hopping is 
\begin{eqnarray}
(j_a,j_b,j_c)=-\frac{ie}{\hbar}&&\left[
       t_{\parallel}\sum_{i,\eta}(\vec{R}_{i+\eta}^{\parallel}-
                                  \vec{R}_{i}^{\parallel})
                          c^{\dag}_{i+\eta}c_i X^{\dag}_{i+\eta}X_i
 	\right. \nonumber \\
    &&+t_{\parallel}\sum_{j,\eta}(\vec{R}_{j+\eta}^{\parallel}-
                                  \vec{R}_{j}^{\parallel})
                          d^{\dag}_{j+\eta}d_j Y^{\dag}_{j+\eta}Y_j
	        \nonumber \\
&& \left.
    +t_{\perp}\sum_{j,\delta}(\vec{R}_{j+\delta}^{\perp}-
                              \vec{R}_{j}^{\perp})
                          d^{\dag}_{j+\delta}c_j Y^{\dag}_{j+\delta}X_j\right],
\nonumber \\
\label{currop:eq}
\end{eqnarray}
The first term corresponds to 
hopping in layer $1$, the second term to hopping in layer $2$, and the 
third to hopping between adjacent positions in the two layers. 
Since we also are going to study thermopower we give the expression for the 
energy current in the same model 
\begin{eqnarray}
(j^e_a,j^e_b,j^e_c)=-\frac{ie}{\hbar}&&\left[
       \frac{t_{\parallel}\epsilon}{2}\sum_{i,\eta}
            (\vec{R}_{i+\eta}^{\parallel}-\vec{R}_{i}^{\parallel})
                          c^{\dag}_{i+\eta}c_i X^{\dag}_{i+\eta}X_i
 	\right. \nonumber \\
    &&+\frac{t_{\parallel}\epsilon}{2}\sum_{j,\eta}
            (\vec{R}_{j+\eta}^{\parallel}-\vec{R}_{j}^{\parallel})
                          d^{\dag}_{j+\eta}d_j Y^{\dag}_{j+\eta}Y_j
	        \nonumber \\
&& \left.
    +\frac{t_{\perp}\epsilon}{2}\sum_{j,\delta}
            (\vec{R}_{j+\delta}^{\perp}-
                              \vec{R}_{j}^{\perp})
                          d^{\dag}_{j+\delta}c_j Y^{\dag}_{j+\delta}X_j\right],
\nonumber \\
\label{Ecurrop:eq}
\end{eqnarray}
and $\epsilon$ is the energy of the quasi-particle. 
Let us separate the current within the layers and perpendicular to the layers, 
since they usually show a different behaviour in experiment. 

\subsection{Current within the layers}
\label{Ill:sect}
In this section we will calculate the current within the layers. 
We split the calculation into two regimes 
because if we use Eq.~(\ref{currop:eq}) 
directly in Eq.~(\ref{kubo:eq}), the result is too complicated to 
decouple, so we split the calculation into low and high temperatures, 
where different parts of the Hamiltonian dominates, and we can use 
perturbation theory. 

\subsubsection{Low Temperatures}
At low temperatures the transport within the layers is coherent. 
If the layers are metallic we can treat them in a Fermi-liquid manner
and use that the conductivity depends on the scattering rate via the 
scattering time, $\tau=h/\Gamma$, 
\begin{equation}
\sigma_{\parallel}=\frac{e^2}{2\pi^2}\int v(\kk)\bar{v}(\kk)
                   \left(-\frac{\partial f}{\partial \epsilon}\right)
                   \tau(\kk)d^2k.
\end{equation}
The decay is calculated as usual from the imaginary part of the self energy,
and is given in Eq.~(\ref{decay:eq}). 
We use the tight binding approximation, Eq.~(\ref{tight:eq}), to get the 
quasiparticle velocity, ${\mathbf v}(\kk)=\frac{\nabla_{\kk}\epsilon}{\hbar}$, 
in one direction and get the conductivity,
\begin{eqnarray}
\sigma_{\parallel}^{xx}&=&\frac{e^2}{\pi h}
\frac{\beta \tilde{t_{\parallel}}a^2}{g^2n_B(1+n_B)} 
\nonumber \\ &\times&
\int_{-2\pi}^{2\pi}dxdy
\frac{\sin^2(x)}
     {1+\cosh[\beta(\epsilon_0+\tilde{t_{\parallel}}\cos(x)+
                               \tilde{t_{\parallel}}\cos(y)-\mu)]}.
\nonumber \\
\label{lowparallel:eq}
\end{eqnarray}

\subsubsection{High Temperatures}
At high temperatures the polarons are localized, the bandwidth disappears, 
and the hopping, $t_{\parallel}$, is the 
perturbation. Utilizing Eq.~(\ref{kubo:eq}) for the current, we decouple the 
electron operators to polaron GFs in each layer, 
$G=(\omega-\Delta+\Sigma+i\Gamma)^{-1}$. 
Note that there is no $\kk$-dependence for the polaron GFs since they are 
localized at an energy 
$\Delta=\epsilon_0-g\hbar\omega_0<0$. 
The calculation of the GF in perturbation theory 
is described in Appendix \ref{SE_high:app}. 
The 4 $X$-operators are decoupled as in Ref.~\onlinecite{mahan} 
into diagonal (no change of boson state) 
and non-diagonal transitions (when the boson-state changes in the hop). 
The result for the non-diagonal transitions is 
\begin{eqnarray}
&&\left\langle T_\tau X^{\dag}(\tau)X(\tau)X^{\dag}(0)X(0)\right\rangle_{\omega}
=\nonumber \\&&
e^{-2g(1+2n_B)}\sum_{l=-\infty}^{\infty}\left\{\int_{-\infty}^{\infty}
 I_l\left[4g\sqrt{n_B(1+n_B)}\right]e^{-l\hbar\omega_0\beta/2}
 e^{il\omega_0t}-1\right\}.
\nonumber \\
\end{eqnarray}
For the diagonal part the four $X$-operators decouple and 
cancels the $-1$ term above when added together.  
Combining the two correlators and taking the imaginary part we convolute the 
two Fourier transforms similarly to what was done for the GF above, 
so that we get for the current: 
\begin{eqnarray}
I_{\parallel}(\omega)&=&\frac{2e}{h}t_{\parallel}^2d^2e^{-2g(1+2n_B)}
\int_{-\infty}^{\infty} \frac{\mathrm{d}\epsilon}{2\pi} A(\epsilon)
\nonumber \\ &\times&
\sum_{l=-\infty}^{\infty}I_l\left[4g\sqrt{n_B(1+n_B)}\right]
e^{-l\hbar\omega_0\beta/2} A(\epsilon+\omega+l\hbar\omega_0)
\nonumber \\ &\times&
\left[n_F(\epsilon)-n_F(\epsilon+\omega+l\hbar\omega_0)\right].
\label{highparallel:eq}
\end{eqnarray}
The  conductance is obtained as usual as 
$\sigma_{\parallel}=e\left.\frac{dI_{\parallel}}{d(\omega)}\right|_{\omega=0}$. 

The conductivity can now be plotted. The metallic, low temperature, 
part decreases with increasing temperature and the insulating, high 
temperature, phase takes over as temperature is increased. There is a peak in 
the resistivity and a crossover from coherent to incoherent transport, 
described by Eq.~(\ref{lowparallel:eq}) and 
Eq.~(\ref{highparallel:eq}) respectively. We did similar plots for a range 
of coupling constants, $g$, and saw that the intralayer crossover occurs 
at a temperature given by
\begin{equation}
k_BT_{\parallel}^{\mathrm{max}} \sim 2\frac{\hbar\omega_0}{g}.
\end{equation}
We have used the same decay, $\Gamma$, for both the low and high-temperature 
limits. This approximation assumes that the dominant part of the 
scattering of the carriers in both limits is the electron-boson coupling. 
The results are shown below in the figures below. 

\subsection{Current perpendicular to the layers}
\label{Iperp:sect}
Let us now turn to the current perpendicular 
to the layers. 
The current operator for an applied field in the perpendicular direction 
in a nearest neighbor hopping model 
(from layer 1 to 2) is given in Eq.~(\ref{currop:eq}). 
We assume that the hopping between the layers only take place between nearest 
neighbors, see Fig.~\ref{layers:fig}. Then, 
$({\bf R}_{j+\delta}-{\bf R}_j)$ is the distance between the two layers, $d$, 
since $\delta=1$ for nearest neighbor hopping. 
The Kubo formula, Eq.~(\ref{kubo:eq}), gives that we have, to second 
order in $t_{\perp}$,
\begin{eqnarray}
I_{\perp}(eV)&=&\frac{2e}{h}t_{\perp}^2d^2\sum_{j,j_1}
               \int_0^{\beta} d\tau e^{ieV\tau}
\left\langle
T_\tau c^{\dag}_j(\tau)d_{j_1}(\tau)
 d^{\dag}_j(0)c_{j_1}(0)\right\rangle
\nonumber \\ &&\times
\left\langle T_\tau Y^{\dag}_{j_1}(\tau)Y_j(\tau)
 X^{\dag}_j(0)X_{j_1}(0)\right\rangle. 
\label{2decop:eq}
\end{eqnarray}
We decouple the operators in the first and second layer respectively. 
This means that the Fourier transformed averages of the electron operators 
gives rise to polaron Green functions
\begin{eqnarray}
&&\left\langle T c^{\dag}_j(t)c_{j_1}(0)\right\rangle \rightarrow 
               G^0_1(\kk,ip_n), \nonumber \\
&&\left\langle T d_{j_1}(t)d^{\dag}_j(0)\right\rangle \rightarrow 
               G^0_2(\pp,ip_n-i\omega). \nonumber
\end{eqnarray}
The average of the polaron-operators ($X,Y$) can be decoupled for the two 
layers separately, and written as an 
exponential $\Phi(t)$,~\cite{lang63,mahan} as done for the GF above. 
Since the coupling in all layers are the same, $M_{1}=M_{2}$, we 
can either combine the two averages of the polaron operators into one 
exponential ($\Phi(t)*\Phi(t)=(\Phi(t))^2$) or keep 
them as two separate. 
%
Then we can perform the Fourier transform $\tau\rightarrow\omega$, 
and if we assume 
that the GFs has an imaginary part, as above, we get the following 
for the interlayer tunneling current 
\begin{widetext}
\begin{eqnarray}
&&I_{\perp}(eV)=\frac{2e}{h}t_{\perp}^2d^2e^{-2g(1+2n_B)}
  \left\{\int_{-\infty}^{\infty}\frac{d\epsilon}{2\pi} 
         \sum_{\kk}A^0_1(\kk,\epsilon)A^0_2(\kk,\epsilon+eV)
        \left[f(\epsilon)-f(\epsilon+eV)\right]\right. \nonumber \\
&&\left. +\left(I_0\left[4g\sqrt{n_B(1+n_B)}\right]-1\right)
          \int_{-\infty}^{\infty}\frac{d\epsilon}{2\pi} 
          \sum_{\kk}A^0_1(\kk,\epsilon)\sum_{\pp}A^0_2(\pp,\epsilon+eV)
          \left[f(\epsilon)-f(\epsilon+eV)\right]\right. \nonumber \\
&&\left. +\sum_{l\neq 0 \atop l=-\infty}^{\infty}
          I_{l}\left[4g\sqrt{n_B(1+n_B)}\right]
          e^{-l\hbar\omega_0\beta/2}
          \int_{-\infty}^{\infty}\frac{d\epsilon}{2\pi} 
          \sum_{\kk}A^0_1(\kk,\epsilon)
          \sum_{\pp}A^0_2(\pp,\epsilon+eV+l\hbar\omega_0)
          \left[f(\epsilon)-f(\epsilon+eV+l\hbar\omega_0)\right]\right\} 
\nonumber \\
\label{current:eq}
\end{eqnarray}
\end{widetext}
$\kk$ belongs to the first layer, and $\pp$ to the second. 
$A^0_1$ and $A^0_2$ are the spectral functions for the electron GFs in each 
layer respectively, 
\begin{eqnarray}
&&A^0_{1}(\kk,\epsilon)=\frac{\Gamma}
                {(\epsilon-\epsilon_{\kk})^2+\Gamma^2}, \\
&&A^0_{2}(\pp,\epsilon)=\frac{\Gamma}
                {(\epsilon-\epsilon_{\pp})^2+\Gamma^2}.
\end{eqnarray}
The index $l$ is a combined index for the number of bosons emitted or 
absorbed in layer 1 and 2 combined

To illustrate this, consider what would be obtained if we had not combined the 
two exponentials, the result would be: 
\begin{widetext}
\begin{eqnarray}
\hspace*{-2cm}I_{\perp}(eV)&\propto&2t_{\perp}^2
e^{-\sum_{\qq}\left(\frac{M_{\qq}}{\hbar\omega_{\qq}}\right)^2(1+2n_B)
   -\sum_{\pp}\left(\frac{M_{\pp}}{\hbar\omega_{\pp}}\right)^2(1+2n_B)}
\prod_{\qq,\pp}\sum_{l,l'=-\infty}^{\infty} \nonumber \\ &&
\times \int_{-\infty}^{\infty}\frac{d\epsilon}{2\pi}
A_2(\epsilon-l'\hbar\omega_{\qq})A_1(\epsilon+eV+l\hbar\omega_{\pp})
\left[n_F(\epsilon-l'\hbar\omega_{\qq})
      -n_F(\epsilon+eV+l\hbar\omega_{\pp})\right] \nonumber \\ &&
\times I_l\left(2\left(\frac{M_{\pp}}{\hbar\omega_{\pp}}\right)^2
         \sqrt{n_B(1+n_B)}\right)
I_{l'}\left(2\left(\frac{M_{\qq}}{\hbar\omega_{\qq}}\right)^2
         \sqrt{n_B(1+n_B)}\right)
e^{-\beta/2(l\hbar\omega_{\pp}+l'\hbar\omega_{\qq})}.
\label{twoaverage:eq}
\end{eqnarray}
\end{widetext}
$l$ belongs to the first layer and counts the number of bosons attached 
to the electron, $l'$ refers in a similar fashion to the second layer. 
The connection between Eq.~(\ref{current:eq}), and Eq.~(\ref{twoaverage:eq}) 
can be found from the identity for the Bessel functions
\begin{equation}
\sum_{l,l'=-\infty}^{\infty}I_{l-l'}(x_1)I_{l'}(x_2)=
\sum_{l=-\infty}^{\infty}I_{l}(x_1+x_2).
\end{equation}

The expression for the current, Eq.~(\ref{current:eq}), has a contribution 
from coherent and two from incoherent transport. 
Note the similarity in structure of Eq.~(\ref{current:eq}) to the expression 
for the GF, Eq.~(\ref{GF:eq}). The first term corresponds to transport 
which conserves the intralayer momentum in the tunneling process. 
This is seen since the crystal momentum $\kk$ is the same for the spectral 
function for the two different layers. 
For the other terms, the intralayer momentum is not conserved, 
(in each layer the sums over the momentum are separate). 
The second row corresponds to transport when the net number of bosons in the 
system is unchanged. When the quasi-particle tunnels it leaves behind the 
cloud of bosons in one layer and attaches to a replica of bosons in the 
second layer. The third row describes transport when a net number of 
bosons is absorbed ($l>0$) or emitted ($l<0$), thus changing the energy of the 
polaron in the hop between the two layers. 
In a recent paper~\cite{lundin04} we established a connection between the 
intralayer coherence and the appearance of dips in the angular 
magnetoresistance, the so called ''magic angles''. We showed that a 
contribution from incoherent jumps between highly conducting one-dimensional 
strands of molecules gives a natural explanation of the phenomena observed in 
the magnetoresistance. 
At low temperature the coherent part dominates but at high temperature 
(high compared to the boson energy, $\hbar\omega_0$, $k_BT>\hbar\omega_0$) 
the incoherent mechanism 
of transport will dominate. Thus, there is a {\em crossover} from coherent to 
incoherent transport. The crossover temperature is fixed by having equal 
contribution from the coherent and the incoherent contributions. Ignoring the 
contribution from the $l\neq 0$ terms in Eq.~(\ref{current:eq}) 
we can get an approximate expression for the crossover temperature as:
\begin{equation}
k_BT^{\mathrm{min}}\sim\frac{\hbar\omega_0}{\sqrt[4]{2^3}g}
\sim 1.68 \frac{\hbar\omega_0}{g}.
\end{equation}

\begin{figure}[hbt]
\includegraphics[width=\columnwidth]{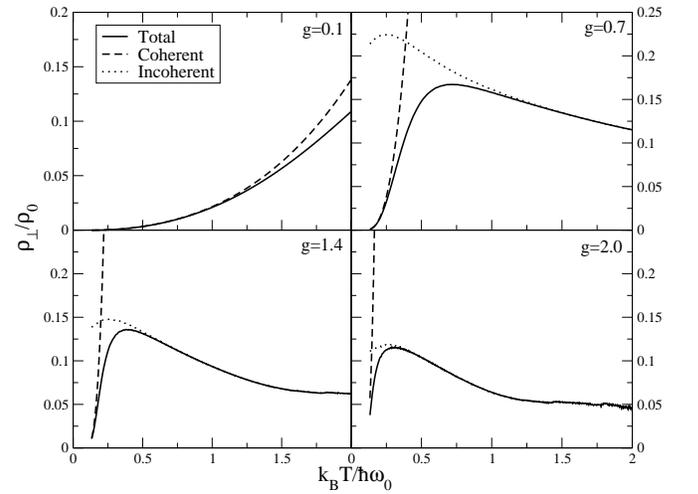}
\caption{\label{crossover:fig}
         Interlayer resistivity as a function of temperature for 
         different values of the coupling $g$ . 
	 At low temperatures, the transport is predominantly 
         coherent, as seen from Eq.~(\ref{current:eq}). Then, as the 
     	 temperature is increased, the incoherent mechanism of transport take 
     	 over. We have 
 	 $\sigma_{\perp}=\sigma_{\perp}^{coh.}+\sigma_{\perp}^{inc.}$, the two 
	 contributions are shown separately and together in the plot. 
         The crossover from coherent to incoherent transport is clearly seen. 
         $\rho_0^{-1}=\frac{2e^2}{h}t_{\perp}^2d^2$, $W=77\hbar\omega_0$.} 
\end{figure}

From Eq.~(\ref{current:eq}) we can extract the conductivity by a simple 
derivative 
$\sigma_{\parallel}=e\left.\frac{dI_{\parallel}}{d(eV)}\right|_{eV=0}$. 
In Fig.~\ref{crossover:fig} we plot the conductivity as a function 
of temperature for one value of $g$. The crossover is clearly seen. 

In general, the interlayer conductivity
for identical decoupled layers is~\cite{moses},
\begin{equation}
\sigma_{\perp}=\frac{2e^2}{h}t_{\perp}^2\int d\epsilon
               \sum_{\kk}|A(\kk,\epsilon)|^2\left[-\frac{df}{d\epsilon}\right],
\label{currcurr2:eq}
\end{equation}
where $A(\kk,\epsilon)$ is the {\em electron} spectral
function for a single layer.
Directly substituting Eq.~(\ref{GF:eq}) in Eq.~(\ref{currcurr2:eq}) we obtain 
the same result found from Eq.~(\ref{current:eq}). 

We can check the result by taking some limits in Eq.~(\ref{current:eq}) 
When the temperature is zero, we get 
\begin{eqnarray}
\sigma_{\perp}(T=0)&=&\frac{2e^2}{h}t_{\perp}^2d^2e^{-2g}
            \sum_{\kk}\frac{A_1(\kk,0)A_2(\kk,0)}{2\pi} \nonumber \\
         &=&\frac{2e^2}{2\pi h}t_{\perp}^2e^{-2g}\frac{D(\epsilon=0)}{2\Gamma}.
\end{eqnarray}
Thus, at low temperature when only the first (coherent) term contributes to the 
conductivity the temperature dependence of $\sigma_{\perp}$ is governed by the 
temperature dependence of the decay, $\Gamma$, 
given by Eq.~(\ref{decay:eq}), for 
the polaron case. 
If we expand Eq.~(\ref{current:eq}) for high temperatures the conductivity 
behaves approximately as: 
\begin{equation}
\sigma_{\perp} \propto T^{-3/2},
\end{equation}
which is consistent with the equipartition theorem~\cite{adler68}. 

If we take the limit $g=0$ we get:
\begin{eqnarray}
\sigma_{\perp}(g=0)&=&\frac{2e^2}{h}t_{\perp}^2d^2
            \sum_{\kk}\int_{-\infty}^{\infty} 
            \frac{d\epsilon}{2\pi}A_1(\kk,\epsilon)A_2(\kk,\epsilon)
\nonumber \\ &&\times
            \beta n_F(\epsilon)\left[1-n_F(\epsilon)\right],
\end{eqnarray}
as expected from transport theory~\cite{mahan}. 

\subsection{Thermopower}
\label{TP:sect}
Let us now turn to calculating the thermopower for intralayer and interlayer 
transport. 
The thermopower is defined as a correlator, using the heat current instead of 
the electrical current in Eq.~(\ref{kubo:eq}), (See Ref.~\onlinecite{mahan}) 
\begin{equation}
L^{12}=\frac{2e}{h}\mathrm{Im}\left\{\int_0^{\beta} dte^{i\omega t}
\langle T \hat{j}^e_{\mu}(t)\hat{j}^{\dag}_{\nu}(0)\rangle\right\},
\label{kuboTP:eq}
\end{equation}
and, using that the current-current correlator gives us
$L^{11}$, and $\sigma=L^{11}/T$, we get 
\begin{equation}
S=\frac{1}{T}\frac{L^{12}}{L^{11}}=\frac{1}{T^2}\frac{L^{12}}{\sigma}.
\label{TP:eq}
\end{equation}
For the intralayer thermopower we consider the low and high temperature limits 
separately. 

\subsubsection{Low temperature intralayer thermopower}
At low temperatures the correlator is similar to the one calculated for the 
intralayer low-temperature conductivity, except for an additional 
$\epsilon_{k}/2$ in the (energy)current operator. This factor only contributes 
an additional $\tilde{t_{\parallel}}\cos(k_xa)/2$ if we assume that we do the 
measurement along $x$. The result is that, for low temperatures, the 
thermopower is
\begin{equation}
S^{\mathrm{low}}_{\parallel}=\frac{1}{T}\frac{\tilde{t_{\parallel}}}{2e}\frac{
\int d^2k f(\epsilon_{kk})\left[1-f(\epsilon_{kk})\right]
          \sin^2(k_xa)\cos(k_xa)}
{\int d^2k f(\epsilon_{kk})\left[1-f(\epsilon_{kk})\right]
          \sin^2(k_xa)}.
\label{Slow:eq}
\end{equation}

\subsubsection{High temperature intralayer thermopower}
At high temperatures, we can follow the same steps as for the intralayer 
current with the only difference that the energy operator is the current 
operator multiplied by $\frac{t_{\parallel}}{2}$. Then, in Eq.~(\ref{TP:eq}) 
the correlators cancel, and we simply end up with~\cite{emin75}:
\begin{equation}
S^{\mathrm{high}}_{\parallel}=\frac{t_{\parallel}}{2eT}.
\label{Shigh:eq}
\end{equation}
The two results in the low and high temperature regions, 
Eq.~(\ref{Slow:eq}) and Eq.~(\ref{Shigh:eq}) respectively, both falls 
off as $\frac{1}{T}$. This means that there would be no peak in the 
intralayer thermopower corresponding to any transition between coherent 
and incoherent transport. Therefore, the transition in the intralayer 
transport is more clearly seen in the electrical transport, not the 
thermopower. The 1/T-dependence is typical for polarons at high 
temperatures as seen, e.g., in  La$_{2/3}$Ca$_{1/3}$MnO$_3$ 
films,~\cite{jaime96} and (La,Ca)MnO$_3$.~\cite{palstra97} 

\subsubsection{Interlayer thermopower}
\begin{figure}
\includegraphics[width=\columnwidth]{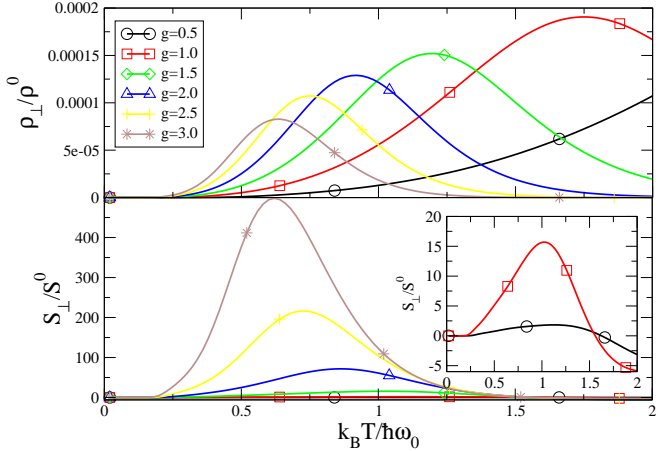}
\caption{\label{TP:fig}
The top panel shows the interlayer resistivity as a function of 
temperature for different electron-phonon coupling strengths, 
$\rho_	0^{-1}=\frac{2e^2}{h}t_{\perp}^2d^2$, $W=160\hbar\omega_0$. 
The peak correspond to the transition between coherent and 
incoherent transport. 
The lower panel shows the interlayer thermopower, $S^0=\frac{t_{\perp}}{2e}$. 
The peak in the thermopower occurs at 
a lower temperature than for the resistivity. 
The inset shows the thermopower for small 
electron-phonon coupling, note that it can change sign although we do not  
consider carriers of hole type here. At high temperatures, and strong 
electron-phonon coupling, the thermopower decays exponentially.}
\end{figure}

We follow the same steps as for the interlayer conductivity with the 
replacement of one current operator by one energy-current operator as 
in Eq.~(\ref{kuboTP:eq}). The result for $L^{12}$ is 
\begin{widetext}
\begin{eqnarray}
&&L^{12}_{\perp}=\frac{2e}{h}t_{\perp}^2e^{-2g(1+2n_B)}d^2
  \left\{\int_{-\infty}^{\infty}\frac{d\epsilon}{2\pi} 
         \sum_{\kk}\xi_{\kk}A^0_1(\kk,\epsilon)A^0_2(\kk,\epsilon+eV)
        \left[f(\epsilon)-f(\epsilon+eV)\right]\right. \nonumber \\
&&\left. +\left(I_0\left[4g\sqrt{n_B(1+n_B)}\right]-1\right)
         \int_{-\infty}^{\infty}\frac{d\epsilon}{2\pi} 
         \sum_{\kk}\xi_{\kk}A^0_1(\kk,\epsilon)\sum_{\pp}A^0_2(\pp,\epsilon+eV)
         \left[f(\epsilon)-f(\epsilon+eV)\right]\right. \nonumber \\
&&\left. +\sum_{l\neq 0 \atop l=-\infty}^{\infty}
         I_{l}\left[4g\sqrt{n_B(1+n_B)}\right]
         e^{-l\hbar\omega_0\beta/2}
         \int_{-\infty}^{\infty}\frac{d\epsilon}{2\pi} 
         \sum_{\kk}\xi_{\kk}A^0_1(\kk,\epsilon)
         \sum_{\pp}A^0_2(\pp,\epsilon+eV+l\hbar\omega_0)
         \left[f(\epsilon)-f(\epsilon+eV+l\hbar\omega_0)\right]\right\}.
\nonumber \\
\label{L12:eq}
\end{eqnarray}
\end{widetext}
Here $\xi_{\kk}=\epsilon_{\kk}-\mu$. 
The thermopower is then given by Eq.~(\ref{TP:eq}). 
In Fig.~\ref{TP:fig} we make a comparative plot of the resistivity 
and the thermopower between the layers. 
For a Fermi-liquid the thermopower would fall off as $\frac{1}{T}$ at 
high temperatures (see, e.g., Salamon {\textit et al.}~\cite{salamon01}), 
fitting a curve to our numerical results shows that our expression for 
the interlayer thermopower falls off \textit{exponentially} instead.  

\subsection{Optical conductivity}
\label{optical:sect}
\begin{figure}[hbt]
\includegraphics[width=\columnwidth]{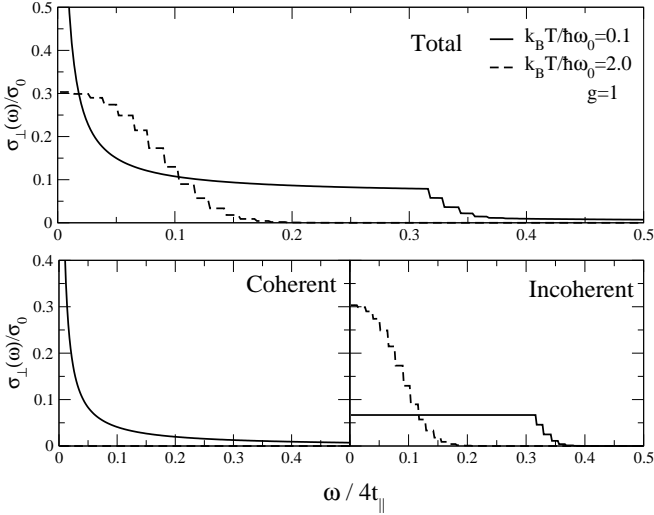}
\caption{\label{opt_cond:fig}
         The top panel shows the frequency dependence of the interlayer 
         optical conductivity for two different 
         temperatures. The Drude peak (at $\omega=0$) 
         disappears when coherence is lost, due to the destruction of 
         coherent quasi-particles with increasing temperature.  
         The two lower panels shows the 
         optical conductivity divided into the two contributions, coherent and 
         incoherent plotted for two different temperatures. The input 
         density of state is flat with a bandwidth of $W=77\hbar\omega_0$.}
\end{figure}

The optical conductivity is given by calculating the derivative of 
the frequency dependent current $I_{\perp}(\omega)$ in  
Eq.~(\ref{current:eq}) with respect to $\omega=eV$. 
Assuming that the following relations hold (when $\Gamma\ll W$) 
\begin{eqnarray}
&&\int dx D(x) A(x,\epsilon)=\int dx D(x) 
       \frac{\Gamma}{(\epsilon-x)^2+\Gamma^2}=
  D(\epsilon) \nonumber \\
&&\int dx D(x) A(x,\epsilon)A(x,\epsilon+\omega)= \nonumber \\
&&\int dx D(x) \frac{\Gamma}{(\epsilon-x)^2+\Gamma^2}
          \frac{\Gamma}{(\epsilon+\omega-x)^2+\Gamma^2}= \nonumber \\
&&D(\epsilon)\frac{\Gamma}{\omega^2+\Gamma^2} \nonumber
\end{eqnarray}
We get that the full expression for the optical conductivity is:
\begin{widetext}
\begin{eqnarray}
&&\sigma(\omega)=\frac{2e^2}{h}t_{\perp}^2d^2e^{-2g(1+2n_B)}
        \left\{\int_{-\infty}^{\infty}\frac{d\epsilon}{2\pi} 
         D(\epsilon)\frac{\Gamma}{\omega^2+\Gamma^2}
         \beta n_F(\epsilon+\omega)\left[1-n_F(\epsilon+\omega)\right]
         \right. \nonumber \\
&&\left. +\left(I_0\left[4g\sqrt{n_B(1+n_B)}\right]-1\right)
          \int_{-\infty}^{\infty}\frac{d\epsilon}{2\pi} 
          D(\epsilon)D(\epsilon+\omega)
          \beta n_F(\epsilon+\omega)\left[1-n_F(\epsilon+\omega)\right]
          \right. \nonumber \\
&&\left. +\sum_{l=-\infty}^{\infty}
          I_{l}\left[4g\sqrt{n_B(1+n_B)}\right]
          e^{-l\hbar\omega_0\beta/2}
          \int_{-\infty}^{\infty}\frac{d\epsilon}{2\pi} 
          D(\epsilon)D(\epsilon+\omega+l\hbar\omega_0)
          \beta n_F(\epsilon+\omega+l\hbar\omega_0)
          \left[1-n_F(\epsilon+\omega+l\hbar\omega_0)\right]
          \right. \nonumber \\
&&\left. +\int_{-\infty}^{\infty}\frac{d\epsilon}{2\pi} 
         D(\epsilon)\frac{2\Gamma\omega}{(\omega^2+\Gamma^2)^2}
         \left[n_F(\epsilon)-n_F(\epsilon+\omega)\right]
         \right. \nonumber \\
&&\left. +\left(I_0\left[4g\sqrt{n_B(1+n_B)}\right]-1\right)
          \int_{-\infty}^{\infty}\frac{d\epsilon}{2\pi} 
          D(\epsilon)D'(\epsilon+\omega)
          \left[n_F(\epsilon)-n_F(\epsilon+\omega)\right]
          \right. \nonumber \\
&&\left. +\sum_{l=-\infty}^{\infty}
          I_{l}\left[4g\sqrt{n_B(1+n_B)}\right]
          e^{-l\hbar\omega_0\beta/2}
          \int_{-\infty}^{\infty}\frac{d\epsilon}{2\pi} 
          D(\epsilon)D'(\epsilon+\omega+l\hbar\omega_0)
          \left[n_F(\epsilon)-n_F(\epsilon+\omega+l\hbar\omega_0)\right]
          \right\}. \nonumber
\end{eqnarray}
\end{widetext}
Here $D'$ is the derivative of the density of states (which is zero for a 
constant DOS). 
In the figure for the optical conductivity, Fig.~\ref{opt_cond:fig}, we see 
that there is a large coherent Drude peak at low frequency, which disappears 
as the temperature increases, consistent with experiments done on the 
manganites~\cite{takenaka99}. 
The disappearance of the Drude peak is consistent with the peak in the 
spectral function disappearing, Fig.~\ref{spectral:fig}. 
Such a behavior has been observed~\cite{katsufuji96} in Sr$_2$RuO$_4$, where 
the Drude peak disappeared above 100K. Another system is 
La$_{0.825}$Sr$_{0.175}$MnO$_4$ where the Drude peak disappears above 
200K.~\cite{takenaka00} 

\subsection{Interlayer Magnetoresistance for a field parallel to the layers}
\label{MR:sect}
We can also make a statement about the magnetoresistance in a certain limit. 
If we apply a magnetic field, $B$, parallel to the layers 
(the x-y plane) there is an orbital effect on the 
paths of the electrons. This can be described by a shift in wave vector,
$\kk\rightarrow\kk-\frac{e}{\hbar}\mathbf{A}$, 
where $\mathbf{A}$ is the vector potential for the magnetic field. 
For a magnetic field in the $x$ direction, when an electron
tunnels between adjacent layers it undergoes a shift in
the y-component of its wave vector by $-d B$~\cite{moses}.
In the general expression Eq.~(\ref{currcurr2:eq}) $|A(\kk,\epsilon)|^2$ 
is replaced with an equation containing 
$A_1(\kk,\epsilon)A_2(\kk + \frac{e}{\hbar} d B \vec{y})$, since there 
will be a difference in the vector potential between the two layers. 

However, since the incoherent part of the conductivity contains a summation 
over $\kk$-space 
and is {\em independent} of $\kk$, this will be unaffected by the 
magnetic field, i.e.\ 
$\sum_{\pp} A_2^0(\pp+\frac{e}{\hbar} d B \vec{y},\epsilon+eV)= 
 \sum_{\pp} A_2^0(\pp,\epsilon+eV)$ since the sum span over the first 
Brillouin zone. 
 
Thus, we will have two contributions to the interlayer
conductivity and one is $B$-independent: 
\begin{equation}
\sigma_{\perp}(B)=\sigma_{\perp}^{\mathrm{coh}}(B)+
                  \sigma_{\perp}^{\mathrm{incoh}}(B=0).
\label{twopart}
\end{equation}
$\sigma^{\mathrm{coh}}(B)$ decreases 
with increasing magnetic field~\cite{schofield,moses}
\begin{equation}
\sigma_{\perp}^{\mathrm{coh}}(B)=\frac{\sigma_{\perp}^{\mathrm{coh}}(B=0)}
             {\sqrt{1+(e v_F c B \Gamma)^2}}.
\end{equation}
where $v_F$ is the Fermi velocity. 
If we increase $B$, the coherent part decreases, and, therefore, 
$T_{\perp}^{\mathrm{max}}$ would shift to {\it lower} values. 
A separation of the conductivity in two parts,
as in Eq.~(\ref{twopart}), has been proposed previously~\cite{hussey98} 
on a phenomenological basis, in order to describe
the magnetoresistance of Sr$_2$RuO$_4$
(Except there a weak field dependence is associated
with the incoherent contribution due to 
Zeeman splitting).

\section{Discussion}
We have presented a layered polaron model for systems consisting of 
two-dimensional layers coupled by tunneling. 
We have found that when the temperature is lower than the characteristic 
boson frequency the physics is dominated by coherent transport where the 
electrons scatters of bosons in the layers. Upon increasing the temperature 
a transition is made into a region where the physics is governed by 
incoherent small polarons. The small polarons are localized at the lattice 
sites and hop to new sites. 
We have extracted results for intralayer and interlayer transport, 
thermopower, ARPES, optical conductivity and magnetoresistance. 

\begin{figure}[hbt]
\includegraphics[width=\columnwidth]{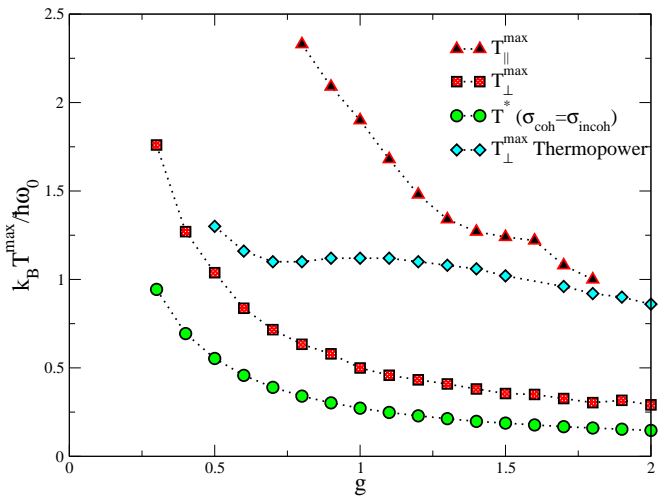}
\caption{\label{T_max:fig} 
         Crossover temperatures as a function of electron-phonon 
         coupling constant, for the intralayer and interlayer resistivity, 
         and the interlayer thermopower, 
         For a fixed coupling and increasing temperature 
         first the coherence between the layers is lost, then there is a 
         crossover peak in the interlayer thermopower, and last the coherence 
         within each layer is lost at elevated temperatures. The sequence 
         of crossover does not change if $t_{\parallel}/W$ is changed.}
\end{figure}

In Fig.~\ref{T_max:fig} we plot the different crossover temperatures 
as a function of the dimensionless electron-boson coupling. 
The different temperature scales associated with small polaron transport is 
summarized in the table~\ref{scales:tab}. 
\begin{widetext}
\begin{table}
\caption{\label{scales:tab}
Temperatures scales obtained in the small polaron model applied to layered 
systems.}
\begin{ruledtabular}
\begin{tabular}{ccccccc}
$\sigma_{\mathrm{coh}}\sim\sigma_{\mathrm{incoh}}$ & & 
GF coherence & & Interlayer transport & & Intralayer transport \\ \hline
$T^*$ & $<$ & 
 $k_BT^{\mathrm{coh}}\sim\frac{\hbar\omega_0}{2g}$ & $\alt$ &
 $k_BT^{\mathrm{max}}_{\perp} \sim \frac{\hbar\omega_0}{\sqrt[4]{2^3}g}$ & $<$&
 $k_BT^{\mathrm{max}}_{\parallel}\sim 2\frac{\hbar\omega_0}{g}$ 
\end{tabular}
\end{ruledtabular}
\end{table}
\end{widetext}

The presented theory differs in the way polarons are formed from the theory 
by Ho and Schofield~\cite{ho02}. In their paper they assume that the polarons 
are formed from phonons traveling along the $c$-direction. Some of the 
results are similar, but in principle it is a different theory. 

\begin{acknowledgments}
U.\ Lundin acknowledges the support from the Swedish foundation for
international cooperation in research and higher education (STINT). 
This work was also supported by the Australian Research Council (ARC). 
\end{acknowledgments}

\appendix

\section{Green function in the high temperature limit}
\label{SE_high:app}
\begin{figure}[hbt]
\includegraphics[width=\columnwidth]{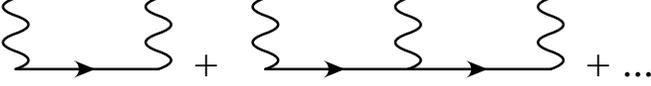}
\caption{\label{sigma:fig}
         The diagrams taken into account when calculating the self-energy 
         for the trapped polaron in the high temperature limit.}
\end{figure}

\begin{figure}[hbt]
\includegraphics[width=\columnwidth]{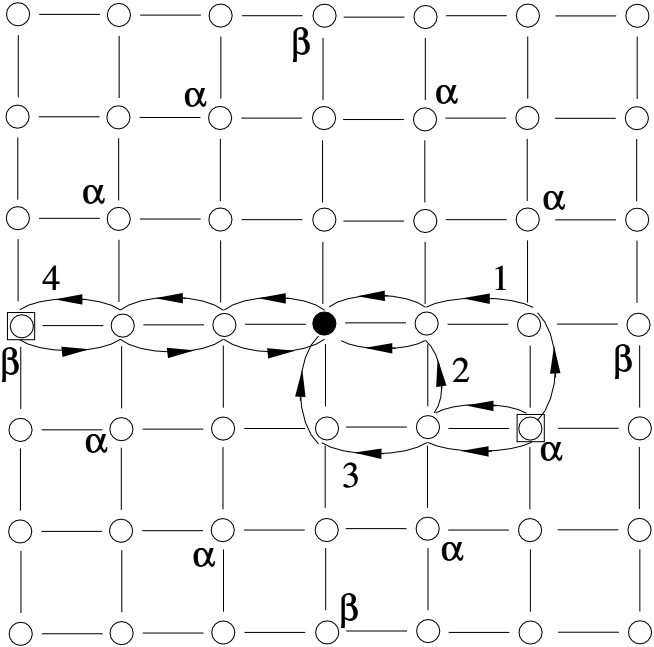}
\caption{\label{2D_paths:fig}
Example when the polaron can hop 3 steps away, from the solid circle 
to the site marked with a box. From the sites marked with
$\alpha$ there are 3 paths back (marked 1-3) and there are 3 paths there,
there are 8 such equivalent $\alpha$-sites.
The polaron can also take the path to the site marked $\beta$, 
there are 4 equivalent sites of this type.}
\end{figure}

In order to calculate the intralayer conductivity at high temperatures, 
when polarons are formed and are localized, we need the GF. 
We perform perturbation theory in $t$ and include the diagrams shown 
in Fig.~\ref{sigma:fig}. 
When summing the series shown in Fig.~\ref{sigma:fig}, we have to consider 
the lattice the electron moves in, i.e.\ 
the number of possible ways to return to the original site. 
In Fig.~\ref{2D_paths:fig} we have shown the number of possible paths 
when the polaron hops 3 jumps away from its original site. 
We have to find a general expression for the number of possible jumps for 
the whole lattice, and take into account identical paths. 
Generally if the electron hops 2n steps, the number of possible paths are
4+2$\cdot$4n(n-1). Thus we get that the self energy is
\begin{eqnarray}
\Sigma(\omega,\Gamma)&=&4\sum_{n=1}^{\infty}[1+2n(1-n)]t^{2n}G^{2n-1}
\nonumber \\
&=& 4t^2G\frac{1+t^2G^2+t^4G^4}{(1-t^2G^2)^3},
\end{eqnarray}
where $G$ is the local polaron GF, and $t$ is the hopping integral within the 
layer. 

\bibliography{paper}

\end{document}